\documentstyle[eqsecnum,aps,twocolumn]{revtex}

\newtheorem{lemma}{Lemma}
\newtheorem{definition}[lemma]{Definition}
\newtheorem{theorem}[lemma]{Theorem}

\newtheorem{corollary}[lemma]{Corollary}

\newenvironment{proof}{{\bf Proof:}}{\hfill$\Box$}
\begin{document}
\draft
\preprint{HEP/123-qed}
\title{Quantum cryptographic three party protocols}
\author{J. M{\"u}ller-Quade and H. Imai}
\address{Imai Laboratory, Institute of Industrial Science, The University of 
Tokyo}
\date{October $31^{st}$, $2000$}
\maketitle
\begin{abstract}
Due to the impossibility results of Mayers and Lo/Chau it is generally
thought that a quantum channel is cryptographically strictly weaker
than oblivious transfer. 

In this paper we prove that in a three party scenario a quantum channel 
can be strictly stronger than oblivious transfer. With the protocol
introduced in this paper we can completely classify the cryptographic
strength of quantum multi party protocols.
\end{abstract}

\section{Introduction}

In a multi party protocol a set $P$ of players wants to correctly
compute a function $f(a_1,\dots,a_n)$ which depends on secret inputs
of $n$ players. Some players might collude to cheat in the protocol as
to obtain information about secret inputs of the other players or
to modify the result of the computation.
Possible collusions of cheaters are modelled by {\em adversary structures}
\begin{definition}
An adversary structure is
a monotone set ${\cal A}\subseteq 2^P$, i.\,e., for subsets
$S'\subseteq S$ of $P$ the property $S\in {\cal A}$ implies $S' \in
{\cal
  A}$.
\end{definition}

The main properties of a multi party protocol are:
{\footnotesize
\begin{enumerate}
\item A multi party protocol is said to be ${\cal A}$-{\em secure} if
no single collusion from $\cal A$ is able to obtain information about
the secret inputs of other participants which cannot be derived from
the result and the inputs of the colluding players.
\item A multi party protocol is ${\cal A}$-{\em partially correct} if no
possible collusion
  can let the protocol terminate with a wrong result.
\item A multi party protocol is called $\cal A$-{\em fair} if no
  collusion from $\cal A$ can reconstruct the result of the multi
  party computation earlier then all honest participants together. No
  collusion should be able to run off with the result.
\end{enumerate}
}
We will be more strict here and demand robustness even against
disruptors. 
{\footnotesize
\begin{enumerate}
\item[2'] A multi party protocol is ${\cal A}$-{\em correct} whenever no
  single collusion from $\cal A$ can abort the protocol, modify its
result, or
  take actions such that some player gets to know a secret value.
\end{enumerate}
}

A protocol is called $\cal A$-{\em robust} if it has all of the above
properties. Note that we will allow only one collusion to cheat, but
we think of every single player as being curious, i.\,e., even if he
is not in the collusion actually cheating he will eavesdrop all
information he can obtain without being detected cheating

With oblivious transfer all multi party protocols can be realized with
perfect security if all players are
cooperating~\cite{BeaGol89,GolLev90,CreGraTap95}. But a collusion of
players can abort the calculation, see next section.

Classically one can avoid this problem only by introducing a new
cryptographic primitive which is more powerful than oblivious
transfer~\cite{ImaMue00STOCS,FitGarMauOst00}. 

This paper analyzes three party quantum protocols and how they can
cope with the problem of disruption. We prove that there are
situations where a quantum channel is strictly more powerful than
oblivious transfer. Together with the results
of~\cite{May96,LoCha96,Lo96} we can conclude that the cryptographic
power of a quantum channel is uncomparable to the power of oblivious
transfer.

\section{Impossibility of classical three party protocols}

To clearly show the advantage of quantum protocols we restate the
following impossibility result of~\cite{ImaMue00Eurocrypt}.

\begin{lemma}\label{MPNoGo}
  Let $P$ be a set of players for which each pair of players is
  connected by a (private) oblivious transfer channel and each player
  has access to an authenticated broadcast channel.  Then $\cal
  A$-robust multi party computations are possible for all functions if
  and only if no two sets of $\cal A$ cover $P\setminus \{ P_i\}$ for
  a player $P_i\in P$ or $|P|=2$.
\end{lemma}

The basic idea to prove this impossibility result
of~\cite{ImaMue00Eurocrypt} is to have two possible collusions
$A_1,A_2$ covering $P\setminus \{ P_i\}$ (for a player $P_i$) where
either all players from $A_1$ or all players from $A_2$ refuse to
cooperate with the players of the other possible collusion. Then the
single player $P_i$ has to assist all other
players. In~\cite{ImaMue00Eurocrypt} it is proven that one cannot
avoid that the player $P_i$ learns a secret.

Especially three party protocols cannot necessarily be realized
robustly if every player is possibly cheating.

\section{Three party protocols}

If all players in a three party protocol cooperate we can use the
protocols of~\cite{ImaMue00QMPforIEICE} to implement $\cal A$-robust
quantum multi party protocols. Hence we focus on the situation where
three players (Alice, Bob, and Helen) want to perform three party
protocols and Alice and Bob are in conflict. One of the two is
refusing to cooperate with the other and it is unclear for Helen who
is cheating.

As a first step we will introduce a bit commitment protocol for Alice
and Bob. The idea is that Alice sends her quantum states via Helen and
Bob does not know which quantum data is
coming from Helen and which data is just forwarded by Helen. Hence Bob
cannot complain without reason or he risks to be detected cheating by
Helen. 
Forwarding information as if it were ones own without being able to
eavesdrop is impossible classically.
One further advantage of the protocol below is that Alice can forward
all information via Helen and Bob cannot know whose information it
is: An anonymous quantum channel. This way Bob cannot distinguish
between commitments of Alice and ``pretended'' commitments of
Helen. Later we want to follow this idea with larger protocols
containing this bit commitment protocol as a subprotocol. Then
we let Alice forward all her information via Helen.

\noindent{\bf Commit}($b$)
{\footnotesize

\noindent{\tt FOR} $i\in\{1,\dots,l\}$ {\tt DO} 
\begin{enumerate}
\item Alice gives a random string $r$ of qubits encoded in random
      bases $s$ $\in \{+,\times\}$ to Helen.
\item Helen sends a substring to Bob
      interleaved with quantum states of her own. 
      Helen
      tells Alice which quantum states are hers without revealing
      information about which substring she forwarded.
\item Bob announces to have received all quantum states.
      With a probability of $1/2$ he publishes all his measurement
      results.  
\item Alice opens to Helen the bases she used. Now Alice is bound to
      the parity bit of $r$.
\end{enumerate}
{\tt OD}
\begin{enumerate}
\item[5] Alice is now bound to the Xor of all quantum states (which
were not measured and published by Bob) Alice announces (via Helen if
needed) if this parity bit is equal to the bit $b$ she originally
wanted to commit to.
\end{enumerate}
}

\noindent{\bf Unveil}
{\footnotesize
\begin{enumerate}
\item Alice opens (via Helen if necessary) all choices she made.
\item Helen and Bob check consistency. 
\end{enumerate}
}

\begin{lemma}
For ${\cal A}=\{ \{Alice\},\{Bob\},\{Helen\}\}$ the above protocol
realizes an ${\cal A}$-robust bit commitment for Alice which binds her
to Bob and to Helen even if Alice and Bob are in conflict.
\end{lemma}

\begin{proof}
We say the protocol has failed if many of the quantum states measured
and published by Bob do not match what Alice sent or what Helen sent. 
If there are only very few cases with discrepancies the protocol is
considered a success.

If the protocol does not fail then the protocol is concealing to
Helen. Helen has sent some substrings to
Bob and as she could not know which substrings would be measured and
published by Bob there are some substrings which she forwarded, but
which were not tested. Hence Helen cannot measure the overall parity
bit $b$ even
after getting to know the bases. Helen cannot measure before getting
to know the bases as she will be detected cheating whenever Bob
measures and publishes a quantum state she disturbed.

If the protocol did not fail it is concealing to Bob unless Helen and
Bob collude. This is clear as Bob does not have the complete quantum
state, and can hence not measure the parity bit.

If the protocol did not fail it is binding for Alice unless she
colludes with Helen, which is impossible according to our assumption.

If the protocol failed there are two cases to be considered. First,
Bob published a lot of measurement results which do not match what
Alice sent, but Helen was not complaining about Bob, then it is clear
for Alice that Helen and Bob collude somehow, but as this is not
possible according to our assumption this will not happen. The second case is
that Helen complains about Bob, then Bob is identified as a cheater as
every player complains about Bob. 
\end{proof}

As Alice is by the above protocol bound to Bob and
Helen we have bit commitment from Alice to Bob and from Alice
to Helen and from Bob to Alice and from Bob to Helen. As the quantum
channel between Alice and Helen and between Bob and Helen is working
we even have oblivious transfer from Helen to Alice and from Helen to
Bob by forcing honest measurements with bit commitment~\cite{Yao95}.

\begin{corollary}\label{CorForcingMeasurments}
The above sketched oblivious transfer protocol between Alice and Helen
and between Bob and Helen is $\cal A$-robust and becomes $\tilde{\cal
A}$-robust after it terminated for ${\cal A}=\{
\{Alice\},\{Bob\},\{Helen\}\}$ and $\tilde{\cal A} = \{
\{Alice,Bob\},\{Helen\}\}$. 
\end{corollary}

\begin{proof}
The bit commitment used for forcing measurments need only be shortly
binding. It need only be binding until the honest measurement is
performed unreversibly. Hence the collusion $\{Alice,Bob\}$, which
would be able to violate the binding condition but not the concealing
condition of the above bit commitment, 
cannot cheat after the measurement is performed.
\end{proof}

Next we have to define some notions which are important for multi
party protocols. Details can be looked up in~\cite{CreGraTap95}.

\begin{definition}
A {\em bit commitment with Xor} ({\em BCX}) to a bit $b$ is a
commitment to bits $b_{1L}$, $b_{2L},\dots,$ $b_{mL},$ $b_{1R},\dots,$
$b_{mR}$
such that for each $i$ $b_{iL}\oplus b_{iR}=b$.
\end{definition}

The following result about zero knowledge proofs on BCX can be found
in~\cite{CreGraTap95} and in references therein.

\begin{theorem}\label{COPYworks}
Bit commitments with Xor allow zero knowledge proofs of linear
relations among several bits a player has committed to using
BCX. Especially (in)equality of bits or a bit string being contained
in a linear code.

Furthermore BCXs can be copied, as proofs may destroy a BCX.
\end{theorem}

In a multi party scenario it is necessary that a player should be
committed to all other players. In our three party case this is given
by our bit commitment protocol which binds one player (Alice or Bob)
to the other two. For Helen the following {\em global bit commitment
with Xor} can be implemented by Corollary~\ref{CorForcingMeasurments}
and the techniques used in~\cite{CreGraTap95} as she is not in
conflict with anyone.

\begin{definition}\label{DefGBCX}
A {\em global bit commitment with Xor} ({\em GBCX}) is a BCX
commitment from a player ${\rm Alice}\in P$ to all other players such
that all players are convinced that Alice did commit to the same bit
in all the different BCX.
\end{definition}

\begin{corollary}
Zero knowledge proofs of linear relations among several GBCX are
possible. Furthermore GBCX can be copied by copying the individual
BCX.
\end{corollary}

On these commitments operates the {\em committed oblivious transfer}
protocol, defined in~\cite{CreGraTap95} which forms the basis of our
multi party protocols.

\begin{definition}
Given two players Alice and Bob where Alice is committed to bits
$b_0,b_1$ and Bob is committed to a bit $a$. Then a {\em committed
oblivious transfer} protocol ({\em COT}) is a protocol where Alice inputs her
knowledge about her two commitments and Bob will input his knowledge
about his commitment and the result will be that Bob is committed to
$b_a$.

In a {\em global committed oblivious transfer} protocol all players
are convinced of the validity of the commitments, i.e., that indeed Bob is
committed to $b_a$ after the protocol.
\end{definition}

As Helen is not actively cheating by assumption and Alice or Bob
cannot complain about Helen without being expelled from the protocol
we can realize GCOT from Helen to Alice and from Helen to Bob by
following the protocol of~\cite{CreGraTap95}.  To realize GCOT from
Alice to Bob is more difficult. We will do this in two steps. First we
realize a subprotocol which we call {\em subGCOT} and second we will
observe that all other steps can be realized easily once one round of
subGCOT was successfull.

To realize subGCOT between Alice and Bob we carry out the first 7
steps of the GCOT protocol of~\cite{CreGraTap95} in a way that Bob
cannot decide if the data comes from Alice or from Helen if he then
complains without reason he risks to get in conflict with Helen, which
would prove him cheating.

{\bf subGCOT}
{\footnotesize
\begin{enumerate}
\item[2] Alice randomly picks $c_0,c_1$ from a previously agreed on
code $\cal C$ (for requirements on $\cal C$ see~\cite{CreGraTap95})
and commits to all bits of the codewords, and proves that the
codewords fulfil the linear relations of $\cal C$ (for the zero
knowledge technique used confer~\cite{CreGraTap95}).
\item[3] Bob randomly picks $I_0,I_1\subset \{1,\dots,M\}$, with
$|I_0|=|I_1| = \sigma m$ ($\sigma$ is a parameter of the code $\cal
C$), $I_1\cap I_0 =\emptyset$ and sets $b^i\leftarrow \overline b$ for
$i\in I_0$ and $b^i \leftarrow b$ for $i\not\in I_0$.
\item[4] Alice runs ${\rm OT}(c_0^i,c_1^i)(b^i)$ with Bob
(by~\cite{Yao95} and the above bit commitment which binds Bob to Helen
and Alice) who gets $w^i$ for $i\in \{1,\dots,m\}$.  Bob tells
$I=I_0\cup I_1$ to Alice who opens $c_0^i,c_1^i$ for each $i\in I$.
\item[5] Bob checks that $w^i = c_{\overline b}^i$ for $i\in I_0$ and
$w^i = c_{b}^i$ for $i\in I_1$, sets $w^i\leftarrow c_b^i$, for $i\in
I_0$ and corrects $w$ using the code $\cal C$'s decoding algorithm,
commits to $w^i$ for $i\in \{1,\dots,m \}$, and proves that $w^1\dots
w^m\in {\cal C}$.
\item[6] All players together randomly pick a subset $I_2\subset
\{1,\dots,m\}$ with $|I_2|=\sigma m$, $I_2\cap I=\emptyset$ and opens
$c_0^i$ and $c_1^i$ for $i\in I_2$.
\item[7] Bob proves that $w^i = c_b^i$ for $i\in I_2$.
\end{enumerate}
} Alice and Helen play subGCOT with Bob in a way that Alice plays via
Helen using the above bit commitment protocol and the
forcing measurement technique of~\cite{Yao95} such that Bob cannot
distinguish between commitments/quantum states of Alice and
pretended commitments/quantum states of Helen, then Bob cannot destinguish
between the subGCOT protocols he plays with Alice and those which are
pretended by Helen. Two cases can occur:
\begin{enumerate}
\item After $l$ trials a subGCOT protocol was successful from Alice to
Bob and neither player complains. This subGCOT protocol can be used to
perform GCOT from Alice to Bob, as all other steps of GCOT are not
critical. 
\item After $l$ trials no subGCOT protocol between Alice and Bob was
successfull. Then, as Helen and Bob do not collude and Bob cannot
distinguish between Alices and Helens data, it is clear that Alice is
cheating if Bob only complained about her data and it is clear that
Bob is cheating if he complained about Helen as well as Alice.
\end{enumerate}

Once Alice and Bob were able to run one round of subGCOT they can
complete this protocol to a GCOT protocol by the steps:
{\footnotesize
\begin{enumerate}
\item[8] Alice randomly picks and announces a privacy amplification
function $h:\{0,1\}^m\rightarrow \{0,1\}$ such that $a_0 = h(c_0)$ and
$a_1 = h(c_1)$ and proves $a_0= h(c_0^1,\dots,c_0^m)$ and $a_1=
h(c_1^1,\dots,c_1^m)$.
\item[9] Bob sets $a\leftarrow h(w)$, commits to $a$ and proves $a =
h(w^1\dots,w^m)$.
\end{enumerate}
}

Alice and Bob give their proofs (following the procedure
of~\cite{CreGraTap95}) in the above steps to Helen. Hence the
proofs must be correct as no one can risk to get into conflict with
Helen, also convincing Helen is enough as she is not colluding with
Alice or with Bob.

For the corectness of the GCOT protocol we refer to the proof
in~\cite{CreGraTap95}. 
We conclude:

\begin{lemma}\label{GCOTworks}
Even if Alice and Bob are in conflict there exists an $\cal A$-robust
protocol for GCOT between Alice and Helen, Bob and Helen and Alice and
Bob. This protocol becomes $\widetilde{\cal A}$-robust after it terminated for ${\cal A}=\{
\{Alice\},\{Bob\},\{Helen\}\}$ and $\tilde{\cal A} = \{
\{Alice,Bob\},\{Helen\}\}$. 
\end{lemma}

\begin{proof}
By Corollary~\ref{CorForcingMeasurments} we can have oblivious
transfer from Helen to Alice and from Helen to Bob. As Helen cannot be
in conflict with Alice or Bob (or a cheater can be identified) we can
realize GCOT from Helen to any other player by the protocols
of~\cite{CreGraTap95} with the security of the oblivious transfer
channel of Corollary\ref{CorForcingMeasurments}.

The above protocol for GCOT between Alice and Bob is still concealing
for Helen even if Alice and Bob collude, but the resulting commitments
need not be binding any more. This is no problem as we allow Alice and
Bob to collude only after the termination of the protocol (See comment
after Theorem~\ref{ThreePart}).
\end{proof}

From a bit commitment which can bind one player to the two other
players we can realize GBCX and together with a GCOT protocol working
in at least one direction between every two players we can obtain all
three party protocols, see~\cite{CreGraTap95}
(and~\cite{ImaMue00Eurocrypt} for creating a {\em distributed bit
commitment}, which is needed in~\cite{CreGraTap95}, in the presence of
conflicts). Hence we can conclude:

\begin{theorem}\label{ThreePart}
For three players Alice, Bob, and Helen all functions can be realized
by quantum multi party protocols if two players are honest. The
protocol becomes $\{ \{Alice\},\{Bob\},\{Helen\}\}\cup \{ A\}$-secure
after its execution if there is a player outside of $A$ nobody
complained about. 
\end{theorem}

Note that the bit commitment used during the protocol is not
necessarily binding after Alice and Bob collude. If one wants to
implement a long binding bit commitment one has to implement it as a
multi party computation.



\section{A completeness Theorem for Quantum Multi Party Protocols}

In the paper~\cite{ImaMue00QMPforIEICE} quantum multi party protocols
were proposed which use secret sharing to force measurements to
implement oblivious transfer. Then these protocols
follow~\cite{ImaMue00Eurocrypt} to implement multi party computations
with oblivious transfer. 

There was an impossibility result in~\cite{ImaMue00QMPforIEICE} that
quantum multi party protocols for all functions become impossible if
two possible collusions cover the set $P$ of players. But due to the
use of oblivious transfer this had to be weakened to the condition of
Lemma~\ref{MPNoGo}. The impossibility result of~\cite{ImaMue00QMPforIEICE} 
did not seem to be sharp. Now we can prove that the result is indeed
sharp as we can implement multi party  protocols even in the case not
covered by~\cite{ImaMue00QMPforIEICE}:

\begin{theorem}
$\cal A$-robust quantum multi party protocols for all functions are
possible if and only if no two collusions of $\cal A$ cover the set
$P$ of players.

These protocols become $\widetilde{\cal A}$-robust after termination
for an adversary structure $\widetilde{\cal A}$ which may contain one
and only one complement of a set of $\cal A$.
\end{theorem}

\begin{proof}
If one looks at~\cite{ImaMue00QMPforIEICE} one can see that the above
theorem is proven there for all cases but one. The case left open is
that two collusions $A_1$, $A_2$, covering $P\setminus \{ P_i\}$ for a
player $P_i$, are in conflict such that no player from $A_1$ can use
the oblivious transfer to any player of $A_2$ (and vice versa).

In this case we can proceed analogously to our three party
protocols. All commitments are made via $P_i$ (Helen) and equality of
commitments is proven to Helen. This way we obtain GBCX in a way that
whenever a player complains a cheater can be identified.

The GCOT protocol can be realized analogously to the above three party
protocol. A player Alice $\in A_1$ runs subGCOT via Helen with a player
Bob $\in A_2$ such that Bob cannot distinguish data from Helen and
data from Alice. Again Bob cannot complain without either being detected
cheating or proving that Alice cheats. On top of subGCOT GCOT can
easily be realized. With GBCX and GCOT we can realize all multi
party protocols~\cite{CreGraTap95} if one keeps in mind that {\em
distributed bit commitments} can be realized without problems even
when conflicts are present~\cite{ImaMue00Eurocrypt}. 

The set $A\in {\cal A}$, for which $A^c$ may be contained in
$\widetilde{\cal A}$, can in the above case be chosen to be any set
containing Helen.
\end{proof}

The set $\widetilde{\cal A}$ can even contain more than one complement
of a set of $\cal A$ provided Helen is not in the additional collusion.
The protocol seems to be more secure than a quantum protocol where no
complains were present. This is true, due to the fact that Helen becomes a
trustable third party in the way that we know she is not colluding with
anyone. This shows that conflicts appearing during the protocol can
yield additional information which can be exploited to increase the
security.


One of the ideas of this paper, namely to anonymize oblivious
transfer, can also be applied to classical protocols. With the
primitive of {anonymous oblivious transfer} all multi party protocols
become possible with perfect security, and whenever a player tries to
abort the protocol this player is identified or the protocol
terminates correctly~\cite{ImaMue00STOCS}.


%
%


\end{document}